\documentclass[preprint,floatfix,prb,showpacs,color,superscriptaddress,epsfig]{revtex4}
\usepackage{graphicx}% Include figure files
\usepackage{dcolumn}% Align table columns on decimal point
\usepackage{bm}% bold math
\usepackage{amsmath,amsfonts,amssymb,amsthm,verbatim}
\usepackage{floatflt}
\usepackage[ansinew]{inputenc}
\usepackage[usenames]{pstcol}

\usepackage{color}
\usepackage{epsfig}

\begin{document}
\title{Transport through a band insulator with Rashba spin-orbit coupling: \\
metal-insulator transition and spin-filtering effects}
\author{T. Jonckheere}
\affiliation{Centre de Physique Th\'eorique, UMR 6207, case 907,
Campus de Luminy, 13288 Marseille Cedex 9, France}
\author{ G.I. Japaridze}
\affiliation{ Andronikashvili Institute of Physics,
Tamarashvili str. 6, 0177 Tbilisi, Georgia}
\author{T. Martin}
\affiliation{Centre de Physique Th\'eorique, UMR 6207, case 907,
Campus de Luminy, 13288 Marseille Cedex 9, France}
\affiliation{Universit\'e de la Mediterran\'ee, Campus de Luminy,
13288 Marseille Cedex 9, France}
\author{R. Hayn}
\affiliation{Institut Mat\'eriaux  Micro\'electronique Nanosciences
de Provence, Facult\'e St.\ J\'er\^ome, Case 142, F-13397 Marseille
Cedex 20, France}
\date{\today}

\begin{abstract}
We calculate the current-voltage characteristic of a one-dimensional
band insulator with magnetic field and Rashba spin-orbit coupling
which is connected to nonmagnetic leads. Without spin-orbit coupling
we find a complete spin-filtering effect, meaning that the electric
transport occurs in one spin channel only. For a large magnetic field which closes the band gap,
we show that spin-orbit coupling leads to a transition from metallic to
insulating behavior. The oscillations of the different spin-components of the
current with the length of the transport channel are studied as
well.
\end{abstract}


%72.25.-b   Spin polarized transport
% 71.70.Ej  Spin–orbit coupling, Zeeman and Stark splitting, Jahn–Teller effect

\maketitle

\section{Introduction}

There is a great interest today to study the phenomena of quantum
transport in low dimensional systems, both from a technological and
a fundamental point of view. Especially important are questions of
spin polarized transport, also known as spintronics. \cite{fert} A
famous example is the proposition of the Datta-Das transistor
\cite{Datta90} which uses the rotation of the electron spin due to
spin-orbit (SO) coupling. There are two sources of spin-orbit
coupling in quasi one-dimensional systems (1D), an intrinsic one due
to the lack of inversion symmetry in certain crystal structures
(Dresselhaus term) \cite{Dresselhaus55} and an external one
triggered by an applied voltage to surface gates (the Rashba SO
coupling). \cite{Rashba60}

Several works studied the SO
coupling and electronic transport in
quasi 1D metallic systems.
\cite{Moroz00,Haeusler01,Seba03,Iucci03,Grivtsev05,Foldi05,Cheng07,Scheid08,Bellucci08,Birkholz09,Japaridze09,RJN_09}
In contrast, the influence of SO coupling and magnetic field
on the transport in 1D band insulators is unexplored, and it can be expected to be
fundamentally different. In the letter band insulators, we will
report on two interesting effects: the
complete spin filtering effect and the SO induced metal-insulator
transition. An incomplete spin filtering effect is possible in 1D
metallic systems with a potential step or additional impurities,
\cite{Seba03,Birkholz09,RJN_09} but the complete spin
filtering as well as the spin-orbit induced metal-insulator
transition which will be reported below are specific to 1D band
insulators and cannot be observed (in principle) in 1D metals.

A prototype model for a one-dimensional (1D) band insulator is a
half-filled ionic chain with alternating on-site energies (energy
difference $\Delta$). Such an ionic chain will be used in our study,
however the obtained results are expected to be generic to any kind
of 1D band insulators, including charge transfer insulators and
realized in diatomic polymers, \cite{AB} as well as the 1D Peierls
insulators, such as polyacetylene. \cite{Su} In a wider sense,
one-dimensional band insulators may also be realized in
carbon-nanotubes. These nanotubes have the advantage that the value
of the gap may be tuned in a very wide range from 600 meV (for
(12,0) nanotubes) up to 8 meV (for (13,0) nanotubes) or even smaller
values. \cite{Hamada92}

Before presenting detailed calculations, let us start with some
qualitative arguments. We first discuss  transport in 1D band
insulators in a magnetic field $B$ and in absence of SO interaction.
Although the magnetic field induced metal-insulator and
insulator-metal transitions have been the subject of studies for
decades, \cite{Brandt_70} in the context of transport in mesoscopic
systems these effects have not been investigated in detail. As we
show in this paper, in the limit of ultra-low temperatures ($T \ll
\Delta$) and strong magnetic field ($ B \geq \Delta $) the field
induced insulator-metal transitions lead to the almost complete spin
filtering effect, since in this case only one spin channel is open
for transport at the Fermi level.

However, the metallic phase reached at $ B \geq \Delta $ shows
unconventional and substantially different properties compared to a
normal metal. As we will show, contrary to the usual 1D metallic
phase, the Rashba spin-orbit coupling opens up a gap again, leading
to a spin-orbit induced metal-insulator transition. It is important
to note that both effects, i.e. the complete spin filtering effect
and the metal-insulator transition induced by the Rashba spin-orbit
coupling are very specific to 1D band insulators, and may not be
observed in 1D metals.

Rather than analyzing the effect of these transitions by computing
the bulk transport properties of the chain, such as the
conductivity, we choose to compute the current of a finite chain of
such a material, whose extremities are connected to metallic
electrodes. A bias is imposed between the electrodes in order to
induce current flow. On the one hand it allows to probe the spin
filtering effects in a setup which is close to experimental
situations, on the other hand it also allows to investigate
potential fluctuations of the current as a function of the chain
length in the presence of SO coupling. In particular, we will show
that a complex behavior, with several periods and a complicated
energy dependence is obtained in the presence of a band gap $\Delta$
and a magnetic field; this is totally different from the simple
harmonic oscillations, with a period inversely proportional to the
SO coupling strength, obtained in the metallic case.

The paper is organized as follows. In Sec. \ref{sec2}, we introduce
the model and in Sec. \ref{sec3} we discuss the spectrum of the
infinite chain. In Sec. \ref{sec4}, we discuss the method which is
used to obtain the  transport properties as well as physical
results. We conclude in Sec. \ref{sec5}.

\section{The model}
\label{sec2}

\begin{figure}
\includegraphics[scale=0.25,angle=0]{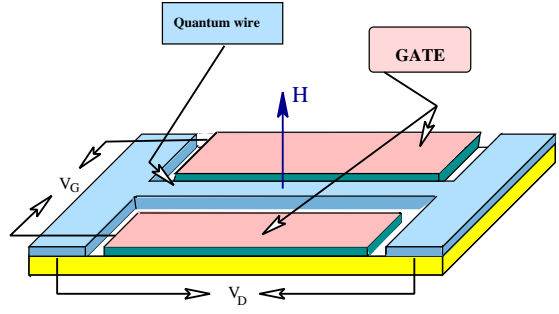}\\%{RI_MIT_Fig1_v3.eps}\\
\caption{(Color online) Schematic figure of the transport process
studied in the paper. The SO coupling parameter $\alpha_R$ is
proportional to $V_G$.} \label{fig1}
\end{figure}

We note first that the spin-orbit coupling can be generated by a
voltage $V_G$ applied to external gates perpendicular to the
current. This is known as Rashba spin-orbit coupling,
\cite{Rashba60} and defines the device studied in the present paper
(Fig.\ \ref{fig1}). We consider a finite chain (oriented in the
$\hat{x}$ direction) connected to metallic leads. Lateral metallic
gates are placed so that to create an electric field which is
perpendicular to both the chain and the magnetic field ($\hat{z}$)
direction.
With these conventions the following Hamiltonian describes the molecular chain:
\begin{eqnarray}
H=&&-t\sum_{n,\sigma}\left(c_{n,\sigma}^{\dagger} c_{n+1,\sigma} +
h.c. \right) + \frac{\Delta}{2} \sum_{n,\sigma} (-1)^n
c_{n,\sigma}^{\dagger} c_{n,\sigma} \nonumber \\ &&- \frac{g \mu_B
H}{2} \sum_{n,\sigma} \sigma  c_{n,\sigma}^{\dagger} c_{n,\sigma}
\nonumber \\ &&+ \alpha_R \sum_n \left( c_{n, \uparrow}^{\dagger}
c_{n+1,\downarrow} - c_{n,\downarrow}^{\dagger} c_{n+1,\uparrow} +
h. c. \right)\; . \label{eq1}
\end{eqnarray}
Here the first contribution describes the kinetic energy in the
tight binding model, the second one accounts for alternating on-site
energies, the third term is the Zeeman coupling (magnetic field $B=g
\mu_B H$) and the last term is the Rashba SO coupling (strength
$\alpha_R$). We consider a finite chain of length $L$ which is
connected to left and right leads by tunneling amplitudes $T_l$ and
$T_r$, respectively. Note that we investigate here the case of {\it
nonmagnetic leads}. We assume that the SO coupling vanishes in the
leads and that the magnetic field only affects the central region
significantly.

\section{The spectrum}
\label{sec3}

\begin{figure}
\includegraphics[scale=0.32,angle=0]{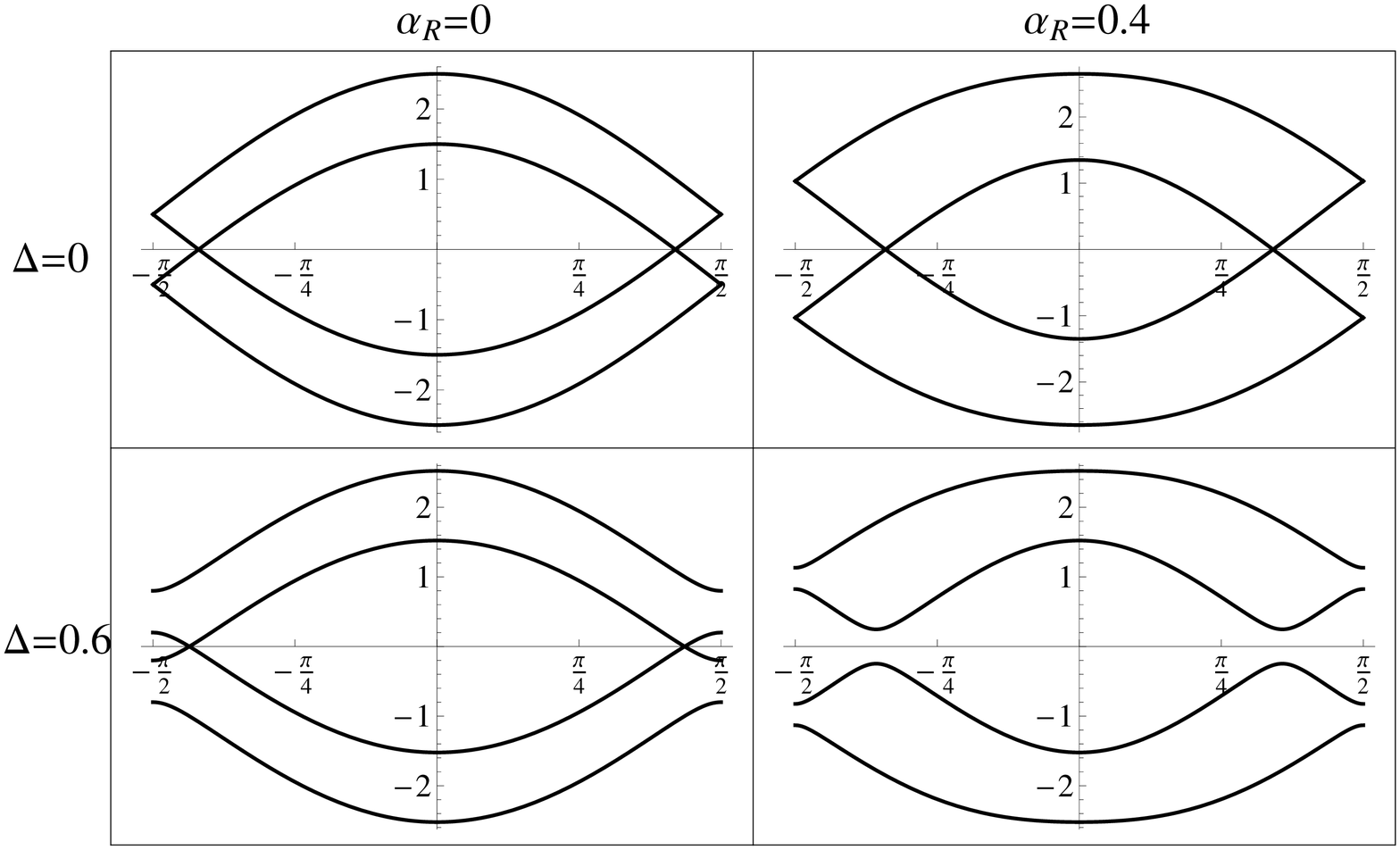}\\
\caption{ Spectrum of the tight-binding chain (see
Eqs.~(\ref{eq1})-(\ref{eq1a})) with magnetic field ($B=g \mu_B
H=1.3$), with and without Rashba coupling $\alpha_R$ and ionicity
$\Delta$ ($t=1$ has been taken as unit of energy).}
\label{fig:spectrum}
\end{figure}

To understand the magneto transport results it is useful to
first consider the spectrum of (\ref{eq1}). For clarity
all spectra are plotted in the reduced Brillouin zone $k\in [-\pi/2a,\pi/2a]$
associated with the presence (possibly small) of alternating on site energies.
Typically this spectrum consists of 4 branches and it
can be obtained exactly:
\begin{eqnarray}
&E^{\pm}_{1/2}(k)=\pm \sqrt{4 \alpha_R^2
 \sin^2 k + \frac{B^2}{4} + \frac{\Delta^2}{4} + 4 t^2 \cos^2 k \pm  W}& \nonumber \\
 &W= \sqrt{16 \alpha_R^2 t^2 \sin^2(2k) + 4 B^2 t^2 \cos^2 k +
 \frac{B^2 \Delta^2}{4}}& \label{eq1a}
\end{eqnarray}
in the general case with spin-orbit coupling $\alpha_R$ and in the
presence of a magnetic field.  It is shown in
Fig.~\ref{fig:spectrum} for different cases of $\Delta$ and
$\alpha_R$, with a non-zero magnetic field $B$.

The upper left corner of Fig. \ref{fig:spectrum} depicts the trivial
case of a non dimerized tight binding chain ($\Delta=0$) in the
presence of a magnetic field. The latter gives rise to a splitting
between the spin up and spin down bands. The spectrum has been
folded in this reduced Brillouin zone to serve as a point of
comparison for the other cases, with dimerisation.

We now consider the case of a non-zero value for $\Delta$ (bottom
left plot of Fig.~\ref{fig:spectrum}). For $\alpha_R=0$, the spin up
and down bands are still separated, but the dimerisation opens a gap
for each spin band at the boundaries of the Brillouin zone. This
implies that for energies close to the Fermi level only one spin
channel will be open for the transport (complete spin filtering
effect, see next Section). As shown on the Figure, the magnetic
field can be so strong that the gap closes and the system can become
metallic. We now switch on the Rashba coupling in the presence of
dimerisation (bottom right corner of~Fig.~\ref{fig:spectrum}). In
this case, the coupling between spin up and spin down gives rise to
an anticrossing, so that the spin-orbit coupling opens up a gap
again.

On the other hand, there is no spin filtering effect for a
homogeneous, metallic chain ($\Delta=0$, top row of Fig.~\ref{fig:spectrum}).
Without magnetic field (not shown), the spin-orbit coupling
can be taken exactly into account by a shift of $k \rightarrow k +
\arctan(\alpha_R/t)$. As can be easily inferred
from the spin split band structure in a magnetic field (left plot) the density of states for spin up and spin
down electrons is the same in that case. And the introduction of spin-orbit coupling (right plot)
does not open a gap. This proves that both effects, i.e.
the complete spin filtering effect and the spin-orbit driven
metal-insulator transition cannot be observed in a metallic system ($\Delta =0$).

\section{Transport through a finite chain}
\label{sec4}

 In the absence of electronic interactions, the current through a finite chain
of length $L$ can be cast exactly in a Landauer type formula,
written here for zero temperature. This current depends on the
orientation of electrons spin at the input lead and the output lead:
the current $I_{s s'}$ for instance, corresponds to electrons which
enter with spin $s$ (with $s =\uparrow \mbox{ or }\downarrow$) from
the left lead and leave the current channel with spin $s'$ to the
right lead. With this convention,
\begin{equation}
I_{ss'}(V_D)= \Gamma_L \Gamma_R \int_{\mu_L}^{\mu_R} \!\!\! dE \; \left| G^{ss'}_{ab}(E) \right|^2
 \; . \label{eq3}
\end{equation}
The  integration is peformed between the chemical potentials of the
left and right leads ($\mu_L=-V_D$ and $\mu_R=0$). The energy
dependent transmission is simply proportional to the square modulus
of the total retarded Green function of the chain (which include the
coupling with the leads) between both endpoints, noted here $a$ and
$b$. The tunneling rates on the left and right side are defined as
$\Gamma_j \equiv 2 \pi \rho_j T_j^2$ ($j=L,R$), where $\rho_j$ is
the (constant) density of states of lead $j$, and $T_j$ the
tunneling amplitude to lead $j$. The total Green function of the
chain between the end sites $a$ and $b$, $G^{ss'}_{ab}$ can be
obtained from the Green function of the bare chain (uncoupled to
leads) $g^{ss'}_{ab}$ by solving the Dyson equations:
\begin{displaymath}
\left( \begin{array}{c}
    G^{\uparrow\uparrow}_{ab} \\ G^{\downarrow\uparrow}_{ab} \\ G^{\uparrow\uparrow}_{bb} \\ G^{\downarrow\uparrow}_{bb} \end{array} \right)
 =
 \left( \begin{array}{c}
    g^{\uparrow\uparrow}_{ab} \\ g^{\downarrow\uparrow}_{ab} \\ g^{\uparrow\uparrow}_{bb} \\ g^{\downarrow\uparrow}_{bb} \end{array} \right)
 +
  \left( \begin{array}{cccc}
  g^{\uparrow\uparrow}_{aa}  & g^{\uparrow\downarrow}_{aa}  & g^{\uparrow\uparrow}_{ab}  & g^{\uparrow\downarrow}_{ab}  \\
  g^{\downarrow\uparrow}_{aa}  & g^{\downarrow\downarrow}_{aa}  & g^{\downarrow\uparrow}_{ab}  & g^{\downarrow\downarrow}_{ab}  \\
  g^{\uparrow\uparrow}_{ba}  & g^{\uparrow\downarrow}_{ba}  & g^{\uparrow\uparrow}_{bb}  & g^{\uparrow\downarrow}_{bb}  \\
  g^{\downarrow\uparrow}_{ba}  & g^{\downarrow\downarrow}_{ba}  & g^{\downarrow\uparrow}_{bb}  & g^{\downarrow\downarrow}_{bb}
  \end{array}  \right)
  \, \left(  \begin{array}{c}
    \Sigma_a G^{\uparrow\uparrow}_{ab} \\ \Sigma_a G^{\downarrow\uparrow}_{ab} \\
     \Sigma_b G^{\uparrow\uparrow}_{bb} \\ \Sigma_b G^{\downarrow\uparrow}_{bb}\end{array} \right)
\end{displaymath}
and similar equations for the opposite spins, and where $\Sigma_j = -i \Gamma_j$ is the retarded self-energy coming from the
coupling to lead $j=L,R$. The Green functions of the bare chain $g^{ss'}_{ab}$ are obtained simply by computing
the eigenvalues and eigenstates of the finite chain, and using a spectral representation:
\begin{equation}
g^{ss'}_{ab}(E)= \sum_n \frac{\psi_n^s(a) \left(\psi_n^{s'}(b)\right)^*}{E-E_n + i 0^+}
\end{equation}
Here all the Green functions, and consequently the current in Eq.~(\ref{eq3}), are 2x2 matrices in
spin space. This is a consequence of the Rashba SO coupling,
which couples the spin-up and spin-down channels. Without SO coupling all quantities
become diagonal in spin space, and the formula for the total Green function reduces to:
\begin{equation}
G^{ss}_{ab} =\frac{g^{ss}_{ab}}{(1 - \Sigma_L g^{ss}_{aa})(1 - \Sigma_R g^{ss}_{bb})
          - \Sigma_L \Sigma_R g^{ss}_{ab} g^{ss}_{ba}}
\end{equation}

\begin{figure}
\includegraphics[width=8.cm]{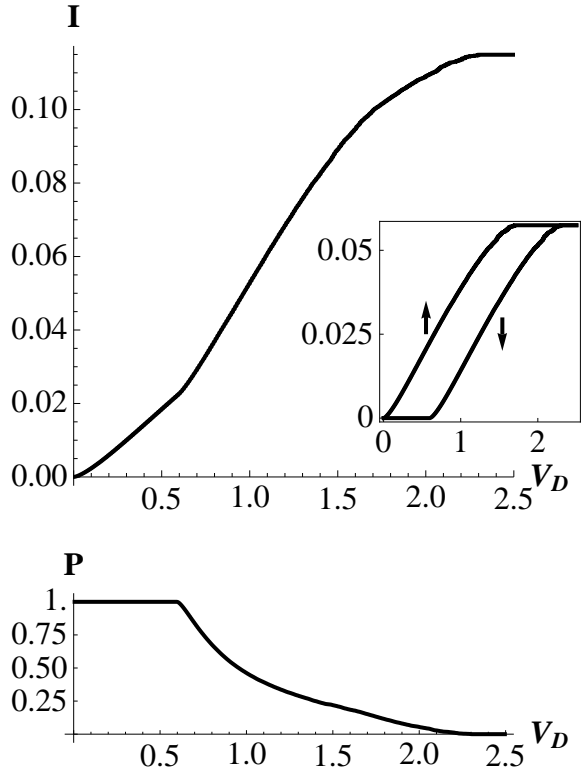}
\caption{Upper plot: total current as a function of the bias voltage
$V_D$, in the {\it spin filtering} configuration. $\Delta=0.6$,
$B=0.6$, $\Gamma_L=\Gamma_R=0.1$ and $t=1$, for a chain of 500
sites.The inset shows the separate contributions from the spin-up
and spin-down current. Lower plot: spin polarization
(Eq.~(\ref{eq4})) for the same parameters.} \label{fig:SF}
\end{figure}

Let us start the discussion of our numerical results with the
current-voltage characteristics in a magnetic field with $\Delta\neq 0$,
but without SO
coupling (see Fig.~\ref{fig:SF}). The magnetic field $B=B_c$ is
chosen such that it just closes the gap, but the exact value of this parameter
is nevertheless not important for the {\em spin-filtering effect}. The transport
for drain voltages between $V_D= 0 $ and $V_D \simeq 0.6 t$ is only possible
for one spin channel. It means that we find complete spin
polarization in the transport channel (connected to nonmagnetic
leads) and a complete spin-filtering. The spin polarization of the current is
defined in the general case as \cite{Seba03, Birkholz09}
%%%%%%%%%%%%%%%%%%%%%%%%%%%%%%%%%%%%%%%%%%%%%%%%%%%%%%%%%%%%%%%%%%%%%
\begin{equation}
P=\frac{I_{\uparrow \uparrow} + I_{\downarrow \uparrow} -
I_{\uparrow \downarrow} - I_{\downarrow \downarrow}}{I_{\uparrow
\uparrow} + I_{\downarrow \uparrow} + I_{\uparrow \downarrow} +
I_{\downarrow \downarrow}} \; .
\label{eq4}
\end{equation}
%%%%%%%%%%%%%%%%%%%%%%%%%%%%%%%%%%%%%%%%%%%%%%%%%%%%%%%%%%%%%%%%%%%%%
As shown on Fig.~\ref{fig:SF},
the spin polarization remains finite (but smaller than unity) for larger voltages
(between approximatively 0.6 $t$ and 2.25 $t$) and disappears at approximatively 2.25 $t$ where the
current reaches saturation (all the electrons of the tight-binding
band contribute). A finite spin polarization means also that the
current creates a total magnetization $M$ in the transport channel
of length $L$. The value of the total magnetization is given by
$M/\mu_B=L(I_{\uparrow \uparrow} + I_{\downarrow \uparrow} -
I_{\uparrow \downarrow} - I_{\downarrow \downarrow}) / \langle v
\rangle$, where $\langle v \rangle$ means the average velocity of
the electrons which are active in the transport process (ballistic
transport).

This spin-filtering effect is expected to work for a wide range of
gap values. The voltage region where only one spin channel is open
is determined by the applied magnetic field. This works also if the
magnetic field is not sufficiently strong to close the gap.
Therefore, even materials with gap values of about 0.5 eV are
possible candidates to show the complete spin-filtering effect. The
onset of the minority spin channel (at zero energy in Fig.\
\ref{fig:SF}) is given by the relative position of the chemical
potential with respect to the upper band edge of the valence band
which may vary from one experimental situation to another.

\begin{figure}
\includegraphics[width=8.cm]{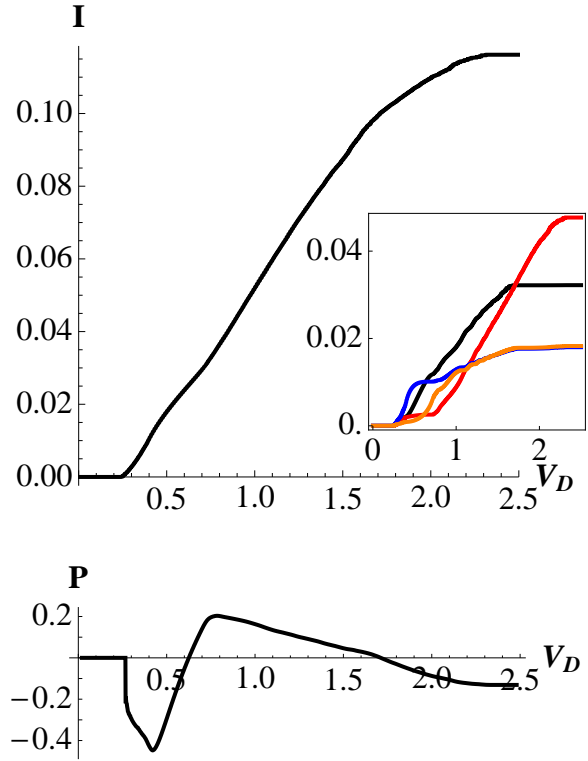}
\caption{(Color online) Upper plot: total current as a function of the bias voltage $V_D$, in the the presence
of Rashba spin-orbit coupling, with $\alpha_R=0.2$, $\Delta=0.6$, $B=0.6$, $\Gamma_L=\Gamma_R=0.1$ and $t=1$
for a chain of 500 sites. The inset shows
the four spin components of the current (in this order from
top to bottom near $V_D=2.5$): $I_{\downarrow \downarrow}$ (red), $I_{\uparrow \uparrow}$ (black),
$I_{\downarrow \uparrow}$ (orange), and $I_{\uparrow \downarrow}$ (blue). Lower plot: spin polarization
(Eq.~(\ref{eq4})) for the same parameters.}
\label{fig:MI}
\end{figure}

We now consider the case of non-zero SO coupling.
The transition from metallic to insulating behavior driven by
SO coupling is shown in Fig.~\ref{fig:MI}. The magnetic
field is the same as in Fig.~\ref{fig:SF}, i.e. it just
closes the gap $B=B_c=\Delta$, and the Rashba SO coupling is
$\alpha_R=0.2 t$. It is created by an external gate
voltage (see Fig.\ \ref{fig1}). The SO coupling leads to an
insulating behavior, as seen in the spectrum (Fig.\ \ref{fig:spectrum}) and
in the current-voltage characteristics (Fig.~\ref{fig:MI}). In
contrast to Fig.~\ref{fig:SF}, the presence of the SO coupling
$\alpha_R$ leads to a current on-set at $V_D \simeq 0.25 t$ corresponding
to half of the gap value for our choice of the chemical potential.
The different current components $I_{ss'}$ are now all different, and
the spin polarization (Eq.~\ref{eq4}) is
different from zero but not complete ($0<P<1$).

\begin{figure}
\centerline{\includegraphics[width=8.5cm]{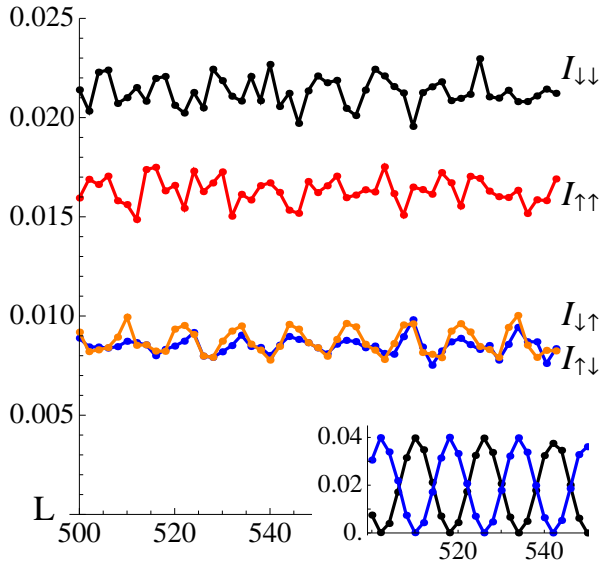}} \caption{(Color
online) Oscillations of the spin components of the current as a
function of the chain length (lengths between 500 and 600), for
$V_D=2.0$,
 when SO Rashba coupling is present
($\alpha_R=0.2$, $\Delta=0.6$, $B=0.6$, $\Gamma_L=\Gamma_R=0.05$ and $t=1$). Inset:
the  same plot with $B=0$ and $\Delta=0$, where $I_{\uparrow\uparrow} =  I_{\downarrow\downarrow}$ and
$I_{\uparrow\downarrow} =  I_{\downarrow\uparrow}$}
\label{fig:oscill}
\end{figure}

Note that the relative values of the different spin-components of the current
in Fig.~\ref{fig:MI} are dependent on the chain length. This is due to the Rashba SO
coupling, which is known to induce spin precession. Here, this spin precession
is made more complex due to the presence of the magnetic field $B$ and the ionicity $\Delta$.
The oscillations of the current components, as a function of the chain length $L$, are shown in
Fig.~\ref{fig:oscill}, for $L$ varying between 500 and 600. These oscillations have a rather
small contrast, show several periods and a complicated dependence on bias voltage
$V_D$ in the general case (a dominating period seems to be present for the off diagonal
components of the current though). This has to be contrasted with the pure
metallic case ($B=0$ and $\Delta=0$, shown in the inset of Fig.~\ref{fig:oscill}),
where only one period $L_p=\pi/\alpha$ is present independently on $V_D$, and
where the contrast is maximum.

\section{Conclusions}
\label{sec5}

In studying the combined effect of magnetic field and SO interaction
on the transport in 1D band insulators we found two interesting
effects. First, already without SO coupling, the presence of a
magnetic field leads to complete spin filtering. We studied this
effect here by connecting the conduction channels to nonmagnetic
leads but the effect of magnetic leads is easy to imagine, at least
qualitatively. Then, spin filtering means high conductance for
parallel magnetization in the leads and low conductance for
antiparallel arrangement.

We speculate that the voltage region of
the spin filtering effect may be dramatically enhanced by
the presence of magnetic
impurities in the band insulator, due to the giant Zeemann effect.
This might be important for the experimental verification
of our proposal.

The second striking effect of this study
appears in band insulators with small band gap
that may be closed by a magnetic field. In that situation, the SO
coupling leads again to an insulating behavior. That is especially
interesting for the Rashba spin orbit coupling which is tuned by a
gate voltage. Therefore, we may propose a device in which the
metal-insulator transition is controlled by the gate voltage via the
Rashba SO term.  This is in sharp contrast with 1D metallic systems, where the SO coupling
does not lead to any metal-insulator transition.

We also showed the oscillations of the different current
components with the chain length. Whereas the simple oscillations in
metallic systems are easy to understand, the oscillations are much
more complex for band insulators. We have let a detailed analysis of
these oscillations for further studies. In our calculations the band
insulator was simulated by an ionic term of alternating on-site
energies in the Hamiltonian. But we think that our results are
generic to any kind of band insulator. On the other hand, the way in
which Coulomb correlations influence our results may be different
from one microscopic Hamiltonian to another. We expect that the
Coulomb correlation just scales the band gap (either to larger or to
smaller values) and that the presented results should remain valid
with effective parameters, however.

The authors thank Marc Bescond and Alvaro Ferraz for useful discussion.

\end{document}